\newcommand{\SiIIa}{Si~{\sc ii} $\lambda 6355$}
\newcommand{\msol}{\mbox{M$_{\odot}$}\ }
\newcommand{\msun}{\mbox{M$_{\odot}$}}

\newcommand{\kms}{\mbox{$\rm{km}\,s^{-1}$}}
\newcommand{\kmsd}{\mbox{$\rm{km}\,\rm{s}^{-1}\,\rm{Mpc}^{-1}$}\ }

\newcommand{\obj}{SN~2022ywc}
\newcommand{\es}{SN~2002es\ }
\newcommand{\bt}{SN~2006bt\ }
\newcommand{\cq}{SN~2007cq\ }

\newcommand{\atg}{iPTF14atg\ }
\newcommand{\jhr}{SN~2016jhr\ }
\newcommand{\yvq}{SN~2019yvq\ }
\newcommand{\cyan}{\textit{cyan}\ }
\newcommand{\orange}{\textit{orange}\ }
\newcommand{\mosfit}{\texttt{MOSFiT}}
\newcommand{\mch}{$M_{\rm Ch}$}
\newcommand{\nhhost}{$N_H^{\rm host}$}

\documentclass[twocolumn]{aastex631}

\shorttitle{SN 2022ywc}
\shortauthors{Srivastav et al.}
\graphicspath{{./}{figures/}}
\begin{document}

\title{Unprecedented early flux excess in the hybrid 02es-like type Ia supernova 2022ywc indicates interaction with circumstellar material}

\correspondingauthor{Shubham Srivastav}
\email{s.srivastav@qub.ac.uk}

\author[0000-0003-4524-6883]{Shubham Srivastav}
\affiliation{Astrophysics Research Centre, School of Mathematics and Physics, Queen's University Belfast, Belfast BT7 1NN, UK}

\author[0000-0001-8385-3727]{T. Moore}
\affiliation{Astrophysics Research Centre, School of Mathematics and Physics, Queen's University Belfast, Belfast BT7 1NN, UK}

\author[0000-0002-2555-3192]{M. Nicholl}
\affiliation{Astrophysics Research Centre, School of Mathematics and Physics, Queen's University Belfast, Belfast BT7 1NN, UK}

\author[0000-0002-0629-8931]{M. R. Magee}
\affiliation{Department of Physics, University of Warwick, Coventry, UK}

\author[0000-0002-8229-1731]{S. J. Smartt}
\affiliation{Department of Physics, University of Oxford, Denys Wilkinson Building, Keble Road, Oxford OX1 3RH, UK}
\affiliation{Astrophysics Research Centre, School of Mathematics and Physics, Queen's University Belfast, Belfast BT7 1NN, UK}

\author[0000-0003-1916-0664]{M. D. Fulton}
\affiliation{Astrophysics Research Centre, School of Mathematics and Physics, Queen's University Belfast, Belfast BT7 1NN, UK}

\author[0000-0002-9774-1192]{S. A. Sim}
\affiliation{Astrophysics Research Centre, School of Mathematics and Physics, Queen's University Belfast, Belfast BT7 1NN, UK}

\author[0009-0005-6989-3198]{J. M. Pollin}
\affiliation{Astrophysics Research Centre, School of Mathematics and Physics, Queen's University Belfast, Belfast BT7 1NN, UK}

\author[0000-0002-1296-6887]{L. Galbany}
\affiliation{Institute of Space Sciences (ICE-CSIC), Campus UAB, Carrer de Can Magrans, s/n, E-08193 Barcelona, Spain}
\affiliation{Institut d’Estudis Espacials de Catalunya (IEEC), E-08034 Barcelona, Spain}

\author[0000-0002-3968-4409]{C. Inserra}
\affiliation{Cardiff Hub for Astrophysics Research and Technology, School of Physics \& Astronomy, Cardiff University, Queens Buildings, The Parade, Cardiff, CF24 3AA, UK}

\author[0000-0001-9598-8821]{A. Kozyreva}
\affiliation{Heidelberger Institut f\"ur Theoretische Studien, Schloss-Wolfsbrunnenweg 35, 69118 Heidelberg, Germany}

\author[0000-0003-1169-1954]{Takashi J. Moriya}
\affiliation{National Astronomical Observatory of Japan, National Institutes of Natural Sciences, 2-21-1 Osawa, Mitaka, Tokyo 181-8588, Japan}
\affiliation{School of Physics and Astronomy, Faculty of Science, Monash University, Clayton, Victoria 3800, Australia}

\author[0000-0002-7975-8185]{F. P. Callan}
\affiliation{Astrophysics Research Centre, School of Mathematics and Physics, Queen's University Belfast, Belfast BT7 1NN, UK}

\author[0000-0002-6527-1368]{X. Sheng}
\affiliation{Astrophysics Research Centre, School of Mathematics and Physics, Queen's University Belfast, Belfast BT7 1NN, UK}

\author[0000-0001-9535-3199]{K. W. Smith}
\affiliation{Astrophysics Research Centre, School of Mathematics and Physics, Queen's University Belfast, Belfast BT7 1NN, UK}

\author[0000-0002-1154-8317]{J. S. Sommer}
\affiliation{Universit\"ats-Sternwarte M\"unchen, Fakult\"at f\"ur  Physik, Ludwig-Maximilians Universität M\"unchen, Scheinerstr. 1, 81679 Munich, Germany}

\author[0000-0003-0227-3451]{J.~P. Anderson}
\affiliation{European Southern Observatory, Alonso de C\'ordova 3107, Casilla 19, Santiago, Chile}
\affiliation{Millennium Institute of Astrophysics MAS, Nuncio Monsenor Sotero Sanz 100, Off. 104, Providencia, Santiago, Chile}

\author[0000-0001-8857-9843]{M. Deckers}
\affiliation{School of Physics, Trinity College Dublin, College Green, Dublin 2, Ireland}

\author[0000-0002-1650-1518]{M. Gromadzki}
\affiliation{Astronomical Observatory, University of Warsaw, Al. Ujazdowskie 4, 00-478, Warszawa, Poland}

\author[0000-0003-3939-7167]{T. E. M\"uller-Bravo}
\affiliation{Institute of Space Sciences (ICE-CSIC), Campus UAB, Carrer de Can Magrans, s/n, E-08193 Barcelona, Spain}
\affiliation{Institut d’Estudis Espacials de Catalunya (IEEC), E-08034 Barcelona, Spain}

\author[0000-0003-0006-0188]{G. Pignata}
\affiliation{Instituto de Alta Investigación, Universidad de Tarapacá, Casilla 7D, Arica, Chile}

\author[0000-0002-4410-5387]{A. Rest}
\affiliation{Space Telescope Science Institute, 3700 San Martin Drive, Baltimore, MD 21218, USA}
\affiliation{Department of Physics and Astronomy, Johns Hopkins University, 3400 North Charles Street, Baltimore, MD 21218, USA}

\author[0000-0002-1229-2499]{D. R. Young}
\affiliation{Astrophysics Research Centre, School of Mathematics and Physics, Queen's University Belfast, Belfast BT7 1NN, UK}



\begin{abstract}

We present optical photometric and spectroscopic observations of the 02es-like type Ia supernova (SN) 2022ywc. The transient occurred in the outskirts of an elliptical host galaxy and showed a striking double-peaked light curve with an early excess feature detected in the ATLAS orange and cyan bands. The early excess is remarkably luminous with an absolute magnitude $\sim -19$, comparable in luminosity to the subsequent radioactively-driven second peak. The spectra resemble the hybrid 02es-like SN 2016jhr, that is considered to be a helium shell detonation candidate. We investigate different physical mechanisms that could power such a prominent early excess and rule out massive helium shell detonation, surface $^{56}$Ni distribution and ejecta-companion interaction. We conclude that SN ejecta interacting with circumstellar material (CSM) is the most viable scenario. Semi-analytical modelling with MOSFiT indicates that SN ejecta interacting with $\sim 0.05$ \msol of CSM at a distance of $\sim 10^{14}$ cm can explain the extraordinary light curve. A double-degenerate scenario may explain the origin of the CSM, either by tidally-stripped material from the secondary white dwarf, or disk-originated matter launched along polar axes following the disruption and accretion of the secondary white dwarf. A non-spherical CSM configuration could suggest that a small fraction of 02es-like events viewed along a favourable line of sight may be expected to display a very conspicuous early excess like \obj.
\end{abstract}



\section{Introduction} \label{sec:intro}

Although supernovae of type Ia (SNe Ia) are valuable as cosmic distance indicators, the nature of their progenitor(s) and explosion mechanism(s) involved are still debated \citep{2018PhR...736....1L,2019NatAs...3..706J}. Modern wide-field surveys %such as Catalina Real-time Transient Survey \citep[CRTS;][]{2009ApJ...696..870D}, Palomar Transient Factory \citep[PTF;][]{2009PASP..121.1334R}, All-sky Automated Search for Supernovae \citep[ASAS-SN;][]{2014ApJ...788...48S}, Panoramic Survey Telescope and Rapid Response System \citep[Pan-STARRS;][]{2016arXiv161205560C}, Zwicky Transient Facility \citep[ZTF;][]{2019PASP..131a8002B}, Asteroid Terrestrial-impact Last Alert System \citep[ATLAS;][]{2018PASP..130f4505T}, Distance Less Than 40 Mpc survey \citep[DLT40;][]{2018ApJ...853...62T}, and Young Supernova Experiment \citep[YSE;][]{2021ApJ...908..143J} 
have unearthed a perplexing diversity within thermonuclear transients, indicating that multiple progenitor systems and underlying explosion mechanisms are likely responsible \citep{2023arXiv230513305L}.

Early-time observations within the first few hours and days from explosion are a sensitive probe for the progenitors and explosion mechanisms of SNe Ia \citep{2018ApJ...861...78M}.
Aided by early discovery and rapid classification, statistical studies of nearby and well-sampled light curves have revealed that a `bump' or `early flux excess' is a characteristic feature in a significant fraction of SNe Ia, $\sim 20\%$ \citep{2022MNRAS.512.1317D,2022MNRAS.513.3035M}. The properties of the early excess are diverse, displaying a range of luminosities, colors and timescales. Different physical mechanisms have been invoked to account for this feature: ejecta-companion interaction within a single-degenerate (SD) scenario \citep{2010ApJ...708.1025K}, radioactive material in the surface layers of SN ejecta \citep{2016ApJ...826...96P,2018ApJ...865..149J,2019ApJ...873...84P,2020A&A...634A..37M}, and interaction with extended circumstellar material or CSM \citep{2021ApJ...909..209P,2023MNRAS.522.6035M}, within either the single or double-degenerate (DD) paradigms.

Although an early excess has been observed in a handful of spectroscopically `normal' SNe Ia like SN 2017cbv \citep{2017ApJ...845L..11H}, SN 2018oh \citep{2019ApJ...870L...1D,2019ApJ...870...12L,2019ApJ...870...13S} and SN 2023bee \citep{2023arXiv230503071H,2023arXiv230503779W}, this feature certainly appears to be more prevalent among members of spectroscopically peculiar Ia sub-classes. A broad, blue bump is seen in a high fraction ($44 \pm 13 \%$) of the shallow silicon subclass of 91T/99aa-like SNe Ia \citep{2018ApJ...865..149J,2022MNRAS.512.1317D}. A short-lived ($\lesssim 1$-day timescale), pulse-like early excess has recently been discovered in the rare carbon-rich, over-luminous subclass of 03fg-like SNe Ia that include SN 2020hvf \citep{2021ApJ...923L...8J}, SN 2021zny \citep{2023MNRAS.521.1162D} and SN 2022ilv \citep{2023ApJ...943L..20S}.

A pronounced excess in blue and ultraviolet (UV) bands that lasts for a few days has been seen in members of the rare subclass of 02es-like SNe Ia, such as iPTF14atg \citep{2015Natur.521..328C} and SN 2019yvq \citep{2020ApJ...898...56M,2021ApJ...919..142B}. SN 2002es \citep{2012ApJ...751..142G} is the prototype of the 02es-like subclass that is characterized by normal-width but sub-luminous light curves, typically displaying peak absolute magnitudes of up to $\sim 1.5$ mag fainter than normal SNe Ia. Spectroscopically, 02es-like events show prominent Ti~{\sc ii} features in their photospheric spectra akin to sub-luminous 91bg-like SNe Ia, with lower ejecta velocities on average relative to normal SNe Ia \citep{2015ApJ...799...52W}. Similar to 91bg-like SNe Ia, 02es-like events show a preference for remote locations within early-type host galaxies \citep{2017hsn..book..317T}.

A subset of 02es-like events display the unusual combination of a high peak luminosity comparable to normal SNe Ia, coupled with Ti~{\sc ii} features in optical spectra around peak, a characteristic associated with cool photospheres of sub-luminous 91bg-like Ia, and have thus been dubbed as `hybrid' 02es-like Ia \citep{2017Natur.550...80J}. \jhr \citep{2017Natur.550...80J} is a well-studied hybrid 02es-like object and showed an early excess that was relatively red. Two other examples in the literature include \bt \citep{2010ApJ...708.1748F} and SN 2007cq \citep{2010ApJS..190..418G,2012MNRAS.425.1789S}, although these two events lack early time photometric coverage to investigate any early excess.

In this paper, we present optical photometric and spectroscopic observations of \obj, a hybrid 02es-like Ia that exhibits a very conspicuous early excess several magnitudes brighter than any previous known examples within different SN Ia sub-classes.

\section{Discovery and Follow-up}\label{sec:discovery}

The Asteroid Terrestrial-impact Last Alert System (ATLAS) is an all-sky survey comprised of $ 4 \times 50\,$cm Schmidt telescopes, each with a $\sim 30$ square degree field of view \citep{2018PASP..130f4505T}. The survey is carried out primarily in two broad filters -- \cyan or $c$-band (roughly equivalent to a composite $g+r$) and \orange or $o$-band (roughly equivalent to a composite $r+i$). A quad of $4 \times 30$ second exposures is obtained for each field of view on a given night, typically reaching $5\sigma$ depths of $\sim 19-19.5$ mag. Difference imaging is performed using the ATLAS all-sky reference catalog \citep[Refcat2;][]{2018ApJ...867..105T}. The data stream is processed through the ATLAS transient science server in order to enable real-time detection and characterization of astrophysical transients \citep{2020PASP..132h5002S}.

AT 2022ywc was discovered by ATLAS on 2022 October 28.13 UT or Modified Julian Date (MJD) 59880.13, at a magnitude of $18.24 \pm 0.06$ in the $c$-band and registered to the Transient Name Server (TNS), designated internally as ATLAS22bkfp \citep{2022TNSTR3139....1T}. ATLAS continued to observe the field, and AT 2022ywc was subsequently identified as a transient of interest on the internal ATLAS transient science server \citep{2020PASP..132h5002S} because of its unusual double-peaked light curve. The transient was classified by the Advanced Public ESO Spectroscopic Survey of Transient Objects or ePESSTO+ \citep{2015A&A...579A..40S} as a 91bg-like SN Ia \citep{2022TNSCR3336....1I} with a spectrum obtained on MJD 59899.29, $\sim 19$ days following the initial discovery.

Final photometry for \obj\ in the $c$ and $o$ bands was obtained using the publicly available ATLAS forced photometry server \citep{2021TNSAN...7....1S}. The individual flux measurements from each nightly quad were combined into a single measurement in order to improve signal to noise, and also to obtain deeper limits in case of non-detections (a factor of $\sqrt{4}$ improvement in signal to noise, or $\sim 0.8$ mag). The stacked forced photometry reveals a $3\sigma$ detection at $m_o = 20.27 \pm 0.26$ on MJD 59878.02, $\sim 2$ days prior to the TNS discovery epoch.

Following the spectroscopic classification of \obj\ as a peculiar SN Ia, and given the unusual nature of the light curve, we obtained further spectroscopic follow-up within ePESSTO+ using the ESO Faint Object Spectrograph and Camera v.2 \citep[EFOSC2;][]{2008Msngr.132...18S} instrument  on the 3.6m New Technology Telescope (NTT). Spectra were obtained on 3 additional epochs with Gr\#13 ($3700-9300\,\mathrm{\AA}$) and the $1\arcsec$ slit, on MJDs 59914.16, 59935.17 and 59957.07. The spectra were reduced and calibrated using the ePESSTO+ data reduction pipeline\footnote{\url{https://github.com/svalenti/pessto}}.

\section{Analysis}

\subsection{Host galaxy, redshift and distance}\label{subsec:host}

\obj\ was discovered in the outskirts of the elliptical galaxy WISEA J032210.70-425303.9. The redshift of the host galaxy, obtained from the NASA Extragalactic Database (NED), is $z=0.061913 \pm 0.000103$ \citep{2009MNRAS.399..683J}. Assuming $H_0 = 70$ \kmsd and a flat Universe with $\Omega_\mathrm{M}=0.3$, the derived luminosity distance is $278$ Mpc, or a distance modulus of 37.22 mag. The SN is offset by $\sim 15\arcsec$ from the galaxy nucleus, corresponding to a projected radial separation of $\sim 21$ kpc at this distance. The Milky Way (MW) extinction in the line of sight is $A_V = 0.04$ mag \citep{2011ApJ...737..103S}. The host extinction is expected to be minimal, given the remote location of the SN within its early-type host. For the analysis that follows, we assume a distance modulus $\mu = 37.22$ mag with a standard systematic uncertainty of 0.15 mag, and a total line-of-sight extinction of $A_V = 0.04$ mag with $R_V = 3.1$.

\subsection{Light curve properties}\label{subsec:lc}

\begin{figure*}
    %\hspace*{-0.8cm}
    \centering
    \includegraphics[width=\linewidth]{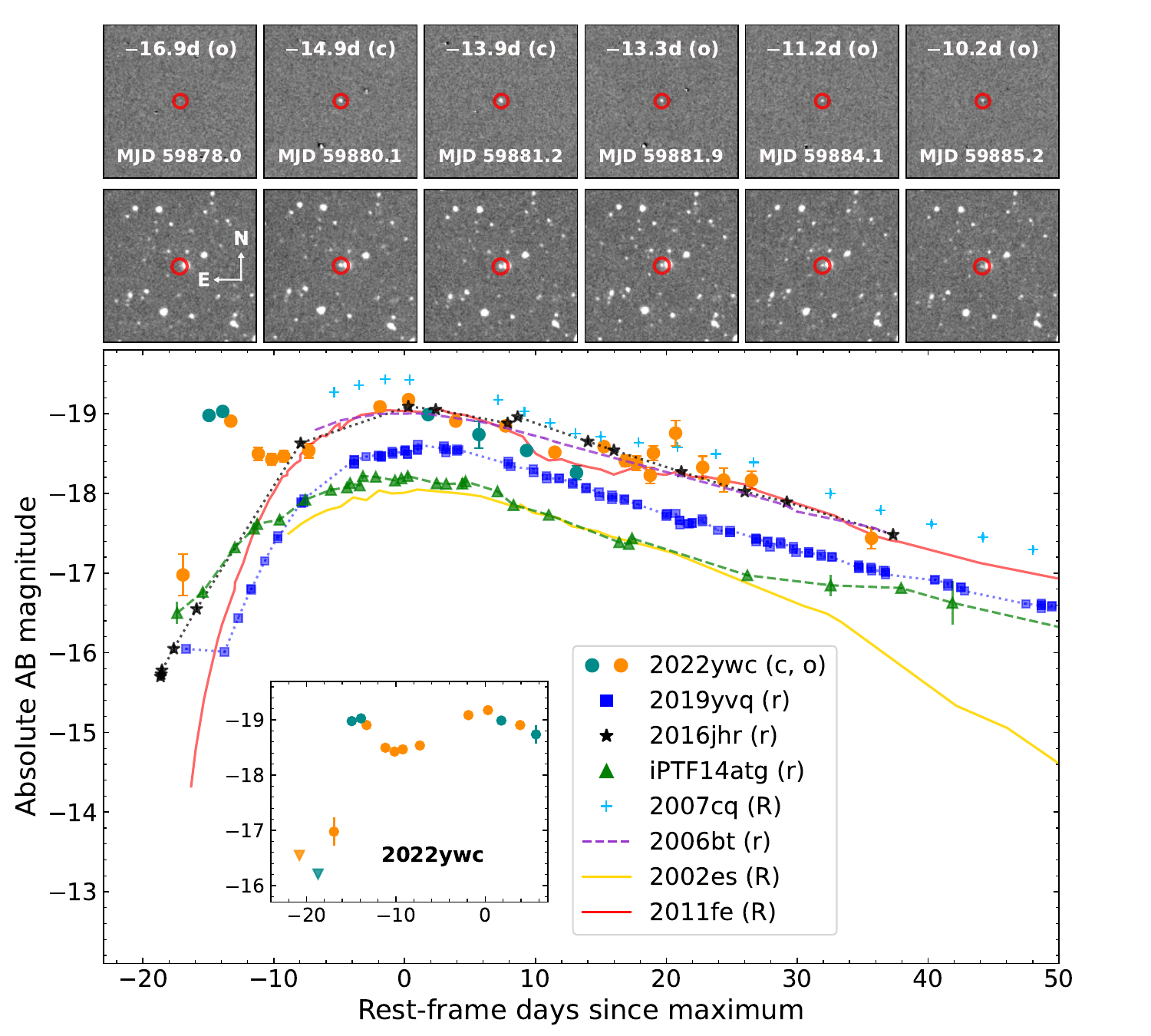}
    \caption{Top and middle panels: Sequence of stacked difference and input images for \obj\ during the early excess. The SN position is marked with a red circle. The image stamps cover a field of view of approximately $6\arcmin \times  6\arcmin$. Bottom panel: Absolute AB magnitude light curve of \obj\ in the ATLAS $c,\,o$ bands, corrected for MW extinction. Also shown for comparison are the MW extinction-corrected light curves of 02es-like events \es \citep{2012ApJ...751..142G}, \bt \citep{2010ApJ...708.1748F}, \cq \citep{2010ApJS..190..418G}, \atg \citep{2015Natur.521..328C}, \jhr \citep{2017Natur.550...80J} and \yvq \citep{2020ApJ...898...56M}, and also the normal Ia SN 2011fe \citep{2016ApJ...820...67Z}. The inset highlights the strikingly luminous early excess for \obj.}
    \label{fig:lc_comp}
\end{figure*}

The extinction-corrected absolute magnitude light curve of \obj\ in the $c$ and $o$ bands is shown in Figure~\ref{fig:lc_comp}. The phase is relative to $o$-band maximum for \obj\ on MJD $59896.0 \pm 2.0$. The epoch of maximum and the peak magnitude in $o$-band was estimated using a third order polynomial fit to the points around peak, excluding the early excess. The uncertainty on the $o$-band peak magnitude was estimated by adding the errors on the measurements around peak in quadrature.
Also shown for comparison are the absolute magnitude $r/R$-band light curves of other 02es-like events \es \citep{2012ApJ...751..142G}, \bt \citep{2010ApJ...708.1748F}, \cq \citep{2010ApJS..190..418G}, \atg \citep{2015Natur.521..328C}, \jhr \citep{2017Natur.550...80J} and \yvq \citep{2020ApJ...898...56M}, along with the normal Ia SN 2011fe \citep{2016ApJ...820...67Z}. $r/R$-band was chosen for a direct comparison since it is the closest to and overlaps with the two ATLAS filters. The literature objects were corrected only for MW extinction, since there is no discernible evidence for significant host extinction. Except for SN 2002es, \bt and \cq where early time photometric coverage is lacking, all the comparison Ia-02es events have a known early excess. The inset in Figure~\ref{fig:lc_comp} shows the early-time light curve of \obj\ in more detail, highlighting the conspicuous early excess with no precedent in the literature.

As shown in Figure~\ref{fig:lc_comp}, 02es-like SNe Ia are typically sub-luminous with peak luminosity $\sim 1$ mag fainter than normal SNe Ia. \obj\ attains a peak luminosity of $M_o \approx -19.2 \pm 0.2$ however, comparable to normal SNe Ia. \obj\ shares this characteristic with at least three other 02es-like events in the literature -- \bt \citep{2010ApJ...708.1748F}, SN 2007cq \citep{2010ApJS..190..418G} and \jhr \citep{2017Natur.550...80J}, together constituting the rare subclass of hybrid 02es-like Ia. Alongside the timescale and luminosity, the color of the early excess is also a useful diagnostic tool for investigating the physical origin of this feature. In the case of \atg \citep{2015Natur.521..328C} and \yvq \citep{2020ApJ...898...56M,2021ApJ...919..142B}, the excess was blue and quite pronounced in the NUV bands, whereas it was relatively red in the case of \jhr \citep{2017Natur.550...80J}. The light curve of \obj\ indicates the early excess was relatively blue, with $(c-o) \equiv (g-i) \lesssim 0$. However, since we only have photometry in the two ATLAS bands, and no concurrent early-time spectroscopic observations, a robust assessment for the color of the excess feature is difficult to make.

\subsection{Light curve fitting}\label{subsec:mosfit}

It is evident that the double-peaked light curve of \obj\ cannot be accounted for by a standard model powered by $^{56}$Ni decay alone. The plausible mechanisms that could potentially explain the luminous early excess are ejecta-companion interaction \citep{2010ApJ...708.1025K}, CSM interaction \citep{2016ApJ...826...96P,2021ApJ...909..209P}, helium (He) shell detonation \citep{2019ApJ...873...84P,2021MNRAS.502.3533M} and surface $^{56}$Ni distribution \citep{2018ApJ...865..149J,2020A&A...634A..37M}. If one of the latter two scenarios were responsible, a very large amount of surface $^{56}$Ni or an extremely thick He shell would be necessary, given the excess feature is comparable in luminosity to the second peak. Copious amounts of iron-group elements (IGEs) and intermediate mass elements (IMEs) in the outer ejecta would in turn induce severe line blanketing at bluer wavelengths, although the spectra (Section~\ref{subsec:spec}) do not show evidence of the same. We discuss the viability and shortcomings of the competing scenarios in more detail in Section~\ref{sec:discussion}. In this Section, we focus on modelling the light curve of \obj\ with a composite model combining $^{56}$Ni decay and CSM interaction. We employ the publicly available Modular Open Source Fitter for Transients \citep[\mosfit;][]{2018ApJS..236....6G,2017ApJ...850...55N}. \mosfit\ takes as input the multi-band photometry of the transient and priors on the parameters for the model being fit to the data. \mosfit\ has a built-in model \textsc{csmni} that combines the luminosity from the decay of radioactive $^{56}$Ni and additional luminosity from CSM interaction, wherein a fraction of the kinetic energy of SN ejecta is converted to radiative energy through collision with dense CSM. The treatment of $^{56}$Ni and $^{56}$Co decay is from \citet{1994ApJS...92..527N}, while the physics of CSM interaction is based on the semi-analytic treatment in \citet{2013ApJ...773...76C}. 

The model is set up such that the contribution from CSM interaction begins at time $t \sim R_0 / v_{\rm ej}$, where $R_0$ is the inner radius of the CSM shell and $v_{\rm ej}$ is the bulk velocity of SN ejecta. The model has 10 free parameters, namely total ejecta mass ($M_{\rm ej}$), $^{56}$Ni mass fraction ($f_{\rm Ni} \equiv M_{\rm Ni} / M_{\rm ej}$), kinetic energy ($E_{\rm k}$), mass of the CSM shell ($M_{\rm CSM}$), inner radius of the CSM shell ($R_0$), CSM density at the initial radius $R_0$ ($\rho_0$), time of explosion relative to first epoch of observation ($t_{\rm exp}$), minimum temperature ($T_{\rm min}$), host galaxy extinction ($A_V^{\rm host}$), and a white-noise variance term ($\sigma$). $T_{\rm min}$ represents the constant temperature the expanding and cooling photosphere settles down to, and $\sigma$ represents the additional uncertainty (in mag) that would make the reduced $\chi^2 = 1$. A power-law density profile for the CSM shell is adopted with $\rho(r) = qr^{-s}$, where the scaling factor $q = \rho_0R_0^s$ \citep{2012ApJ...746..121C}. The power-law index was fixed to $s = 2$ corresponding to a steady-wind CSM model \citep[eg.][]{2011ApJ...729L...6C}. The $\gamma$-ray opacity was fixed at $\kappa_{\gamma}=0.027$ cm$^2$ g$^{-1}$ \citep{1997A&A...328..203C}. An average photospheric velocity for the SN ejecta is inferred from the free parameters $M_{\rm ej}$ and $E_{\rm k}$ assuming a constant density \citep{1982ApJ...253..785A}, using 
$$E_{\rm k} \approx \frac{3}{10} M_{\rm ej} v_{\rm ej}^2$$

\begin{figure*}
    %\hspace*{-0.8cm}
    \centering
    \includegraphics[width=0.9\linewidth]{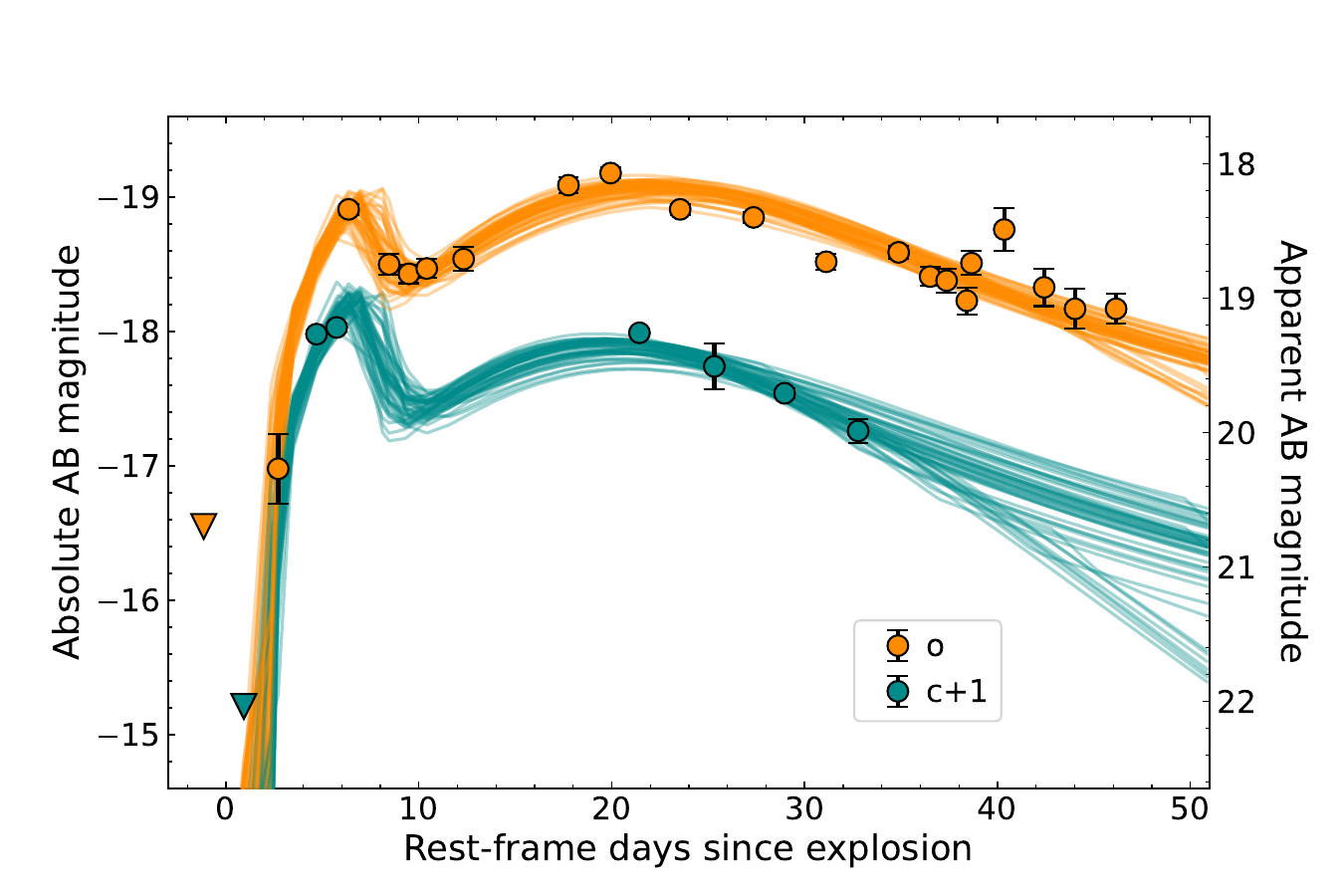}
    \caption{ATLAS light curve of \obj\ along with draws from the posterior of the \textsc{csmni} model from \mosfit. The $c$-band light curve and model is shifted for clarity. The explosion epoch, which is one of the free parameters in the model, is MJD $59875.1^{+0.6}_{-0.7}$.}
    \label{fig:csmni}
\end{figure*}

We use the dynamic nested sampling approach within \mosfit\ implemented using the \textsc{dynesty} package \citep{2020MNRAS.493.3132S} to evaluate the posterior distributions of the model parameters. Broad priors were assigned for the free parameters, with $M_{\rm ej} \in [0.1, \,2.0]$ \msol to allow for a range of WD progenitor masses and $E_{\rm k} \in [0.1, \,2.0]$ $\times 10^{51}$ erg, informed by constraints on the ejecta velocity from spectroscopic observations. Since we have a good constraint on the explosion epoch from high cadence pre-discovery non-detections (Figure~\ref{fig:csmni}), a narrow prior of $t_{\rm exp} \in [-4, 0]$ days was set, where $t_{\rm exp}$ is defined relative to the first photometric observation on MJD 59878.02. We use log-flat priors for parameters with a range that spans 2 or more orders of magnitude in order to allow for a more efficient and robust exploration of the parameter space.

\begin{figure*}
    %\hspace*{-0.8cm}
    \centering
    \includegraphics[width=0.95\linewidth]{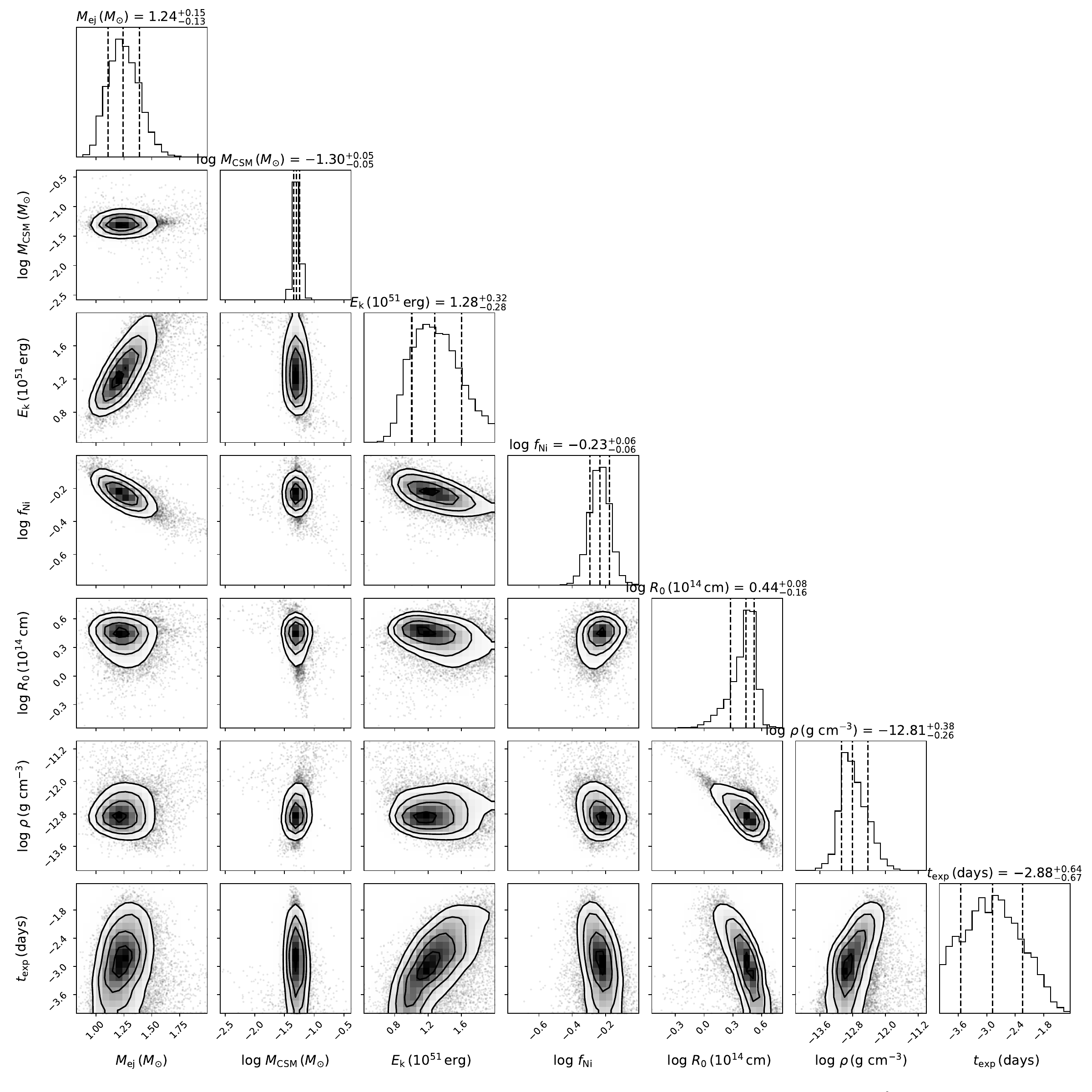}
    \caption{2D posteriors for the main physical parameters in the \mosfit\ \textsc{csmni} model fit to \obj. $R_0$ is in units of $10^{14}$ cm and $t_{\rm exp}$ is relative to MJD 59878.02.}
    \label{fig:corner}
\end{figure*}

Table~\ref{tab:params} summarizes the priors and the best-fit median values and $1\sigma$ bounds for the free parameters in the \textsc{csmni} model for \obj. The light curve fit is shown in Figure~\ref{fig:csmni}, and Figure~\ref{fig:corner} shows a corner plot with two-dimensional posteriors for the model parameters.

The \textsc{csmni} model reproduces the timescale, luminosity and color of the early excess and also the second peak and overall light curve shape fairly well. The most interesting physical parameters in the model are the explosion parameters $M_{\rm ej}$, $M_{\rm Ni}$ and $E_{\rm k}$, and the key CSM parameters $M_{\rm CSM}$ and $R_0$. The key parameters are well-constrained, as seen in Figure~\ref{fig:corner}. $M_{\rm ej} = 1.24^{+0.15}_{-0.13}$ \msol is consistent with either a near Chandrasekhar-mass (\mch) or \mch\ WD progenitor within the uncertainties. The $^{56}$Ni fraction $f_{\rm Ni}$ implies $M_{\rm Ni} = 0.73^{+0.21}_{-0.17}$ \msun, compatible with normal SN Ia luminosity.
%Power law index s: -2 (wind-like CSM density profile). s can also be a free parameter but difficult to constrain given fitting only 2 bands, and CSM is likely aspherical.
An average ejecta velocity, inferred from the median $M_{\rm ej}$ and $E_{\rm k}$ from the fit, is $v_{\rm ej} \sim 13000$ \kms. This is comparable to the \SiIIa\ velocity measured for \obj\ (Section~\ref{subsec:spec}), and consistent with the photospheric velocity of hybrid 02es-like Ia \bt \citep{2010ApJ...708.1748F} and \jhr \citep{2017Natur.550...80J} around peak. The model requires $\sim 0.05$ \msol of CSM at $\sim 3 \times 10^{14}$ cm from the WD progenitor in order to explain the early excess. We note here that the \citet{2013ApJ...773...76C} model assumes optically thick interaction and may therefore be unreliable at low CSM masses. The model favors a minimal host extinction as expected, and the estimated time of explosion is MJD $59875.1^{+0.6}_{-0.7}$.

\begin{table*}
\centering
\caption{Summary of priors and median values with 16th/84th percentiles of the marginalized posteriors for the \textsc{csmni} model parameters.}
\begin{tabular}{ccccc}
\hline
parameter & units & prior & prior type & best-fit\\
\hline
$M_{\rm ej}$ & \msun & [0.1, 2] & flat & $1.24^{+0.15}_{-0.13}$\\
$f_{\rm Ni}$ & dimensionless & [0.01, 1] & log-flat & $0.59^{+0.09}_{-0.08}$\\
$E_{\rm k}$ & $10^{51}$ erg & [0.1, 2] & flat & $1.28^{+0.32}_{-0.28}$\\
$M_{\rm CSM}$ & \msun & [0.001, 1] & log-flat & $0.05^{+0.01}_{-0.01}$\\
$R_0$ & $10^{14}$ cm & [0.015, 15] & log-flat& $2.73^{+0.58}_{-0.84}$\\
$\rho_0$ & $10^{-13}$ g cm$^{-3}$ & [0.01, 100] & log-flat & $1.55^{+2.17}_{-0.70}$\\
$t_{\rm exp}$ & days & [-4, 0] & flat & $-2.88^{+0.64}_{-0.67}$\\
$T_{\rm min}$ & $10^3$ K & [0.1, 100] & log-flat & $5.89^{+1.42}_{-1.36}$\\
$A_V^{\rm host}\dag$ & mag & [0.0, 5.6] & log-flat & $0.002^{+0.036}_{-0.002}$\\
\hline
\end{tabular}
\label{tab:params}
%\tablenotemark{a}
\tablenotetext{}{$\dag$ The free parameter in the model is the hydrogen column density along the line of sight within the host galaxy, \nhhost. We use a prior of $N_H^{\rm host} \in [10^{16},\,10^{22}]$ cm$^{-2}$. The host extinction is computed from the column density using the relation $A_V^{\rm host} \approx N_H^{\rm host} / (1.8 \times 10^{21})$ mag \citep{1995A&A...293..889P}.}  
\end{table*}

\subsection{Spectroscopic properties}\label{subsec:spec}

\begin{figure*}
    \centering
    \includegraphics[width=0.8\linewidth]{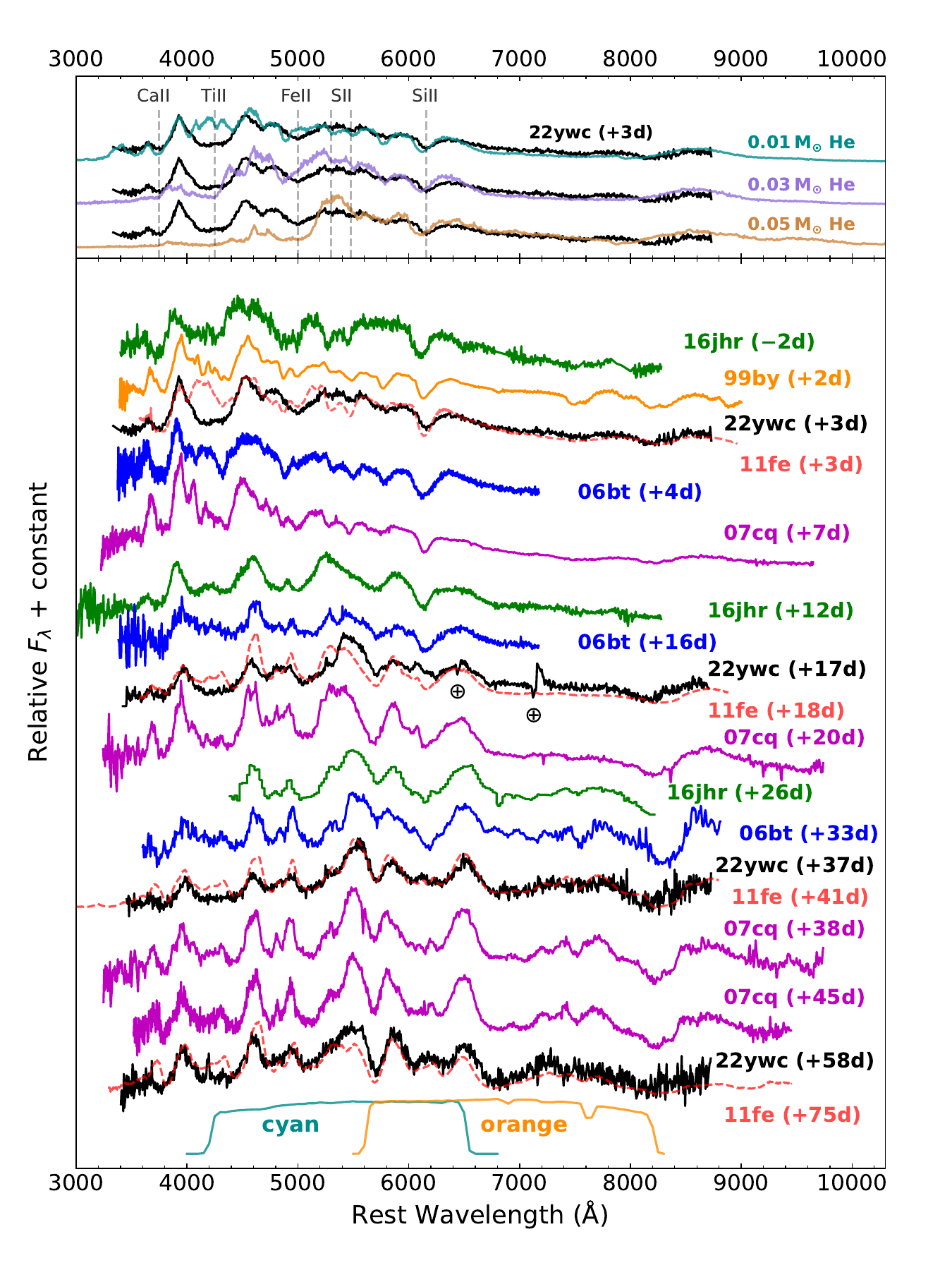}
    \caption{Spectral evolution of \obj\ between +3d and +58d relative to maximum. Also plotted for comparison are the spectra of the hybrid 02es-like Ia \bt \citep{2010ApJ...708.1748F}, \cq \citep{2010ApJS..190..418G} and \jhr \citep{2017Natur.550...80J}, the 91bg-like SN 1999by \citep{2008AJ....135.1598M} and the normal Ia SN 2011fe \citep{2014MNRAS.439.1959M,2016ApJ...820...67Z} at similar phases. Top panel shows the $+3$d spectrum of \obj\ compared to three double detonation models of \citet{2019ApJ...873...84P}, computed for a WD mass of $1.0$ \msol and He shell masses of 0.01, 0.03 and 0.05 \msun. The model spectra were shifted along y-axis for clarity, but not scaled in wavelength or flux axes relative to the observed spectrum. Prominent telluric features are marked with the symbol $\oplus$. The ATLAS \orange and \cyan filter response curves are also shown for reference.}
    \label{fig:spectra}
\end{figure*}

The four NTT spectra obtained for \obj\ at rest-frame phases +3d, +17d, +37d and +58d relative to the time of maximum (MJD 59896.0) are shown in Figure~\ref{fig:spectra}. Also shown for comparison are spectra of the hybrid 02es-like events \jhr \citep{2017Natur.550...80J}, \cq \citep{2010ApJS..190..418G} and \bt \citep{2010ApJ...708.1748F,2012MNRAS.425.1789S}, the 91bg-like SN 1999by \citep{2008AJ....135.1598M} and also the normal Ia SN 2011fe \citep{2014MNRAS.439.1959M,2016ApJ...820...67Z}. The spectra were corrected for MW extinction \citep{2011ApJ...737..103S} and Doppler-corrected to the rest-frame. The first spectrum for \obj\ at +3d exhibits a fairly blue continuum similar to the comparison SNe, and in contrast to the likes of massive He shell detonation candidates SN 2020jgb \citep{2023ApJ...946...83L}, SN 2018byg \citep{2019ApJ...873L..18D} and SN 2016dsg \citep{2022ApJ...934..102D}, that show a red continuum blue-ward of $\sim 5000\,{\rm \AA}$ in spectra around peak due to substantial line-blanketing from IMEs and IGEs synthesized within the He ash.

From the first spectrum of \obj\ at $+3$d, we measure a \SiIIa\ velocity of $\sim 9800 \pm 400$ \kms. The ``W" feature $\sim 5500\,$\AA\ attributed to S~{\sc ii} is noticeably weaker in \obj\ when compared to SN 2016jhr, SN 2006bt and SN 2011fe. The most conspicuous feature in the +3d spectrum of \obj\ is the broad absorption blend $\sim 4500\,$\AA, attributable to Ti~{\sc ii}. The presence of Ti~{\sc ii} features in the spectra has been suggested as evidence for products of He shell burning in the double detonation scenario \citep[eg.][]{2010ApJ...719.1067K,2019ApJ...873...84P, 2022MNRAS.517.5289C}. 02es-like SNe Ia are often classified as 91bg-like owing to the presence of strong Ti~{\sc ii} features around $4000-5000\,$\AA. However, there are subtle differences, with 91bg-like Ia showing stronger O~{\sc i} and Ca~{\sc ii} features. These differences, in conjunction with the distinct photometric behaviour (i.e. slower decline rate and higher luminosity for 02es-like Ia), can be used to tell these sub-classes apart.

The spectra of \obj\ show a good overall match with the hybrid 02es-like \bt \citep{2010ApJ...708.1748F}, \cq \citep{2010ApJS..190..418G} and \jhr \citep{2017Natur.550...80J}. The early spectra show marked differences with the normal Ia SN 2011fe, especially at bluer wavelengths. However, later spectra at +37d and +58d are much closer to SN 2011fe.
The $+3$d spectrum of \obj\ does not show any obvious signatures of He although this does not necessarily rule out its presence in the ejecta, since formation of He lines requires non-thermal excitation \citep[eg.][]{2012MNRAS.422...70H}. Moreover, simulations show that He lines may be easier to detect at near infrared (NIR) wavelengths, specifically the He~{\sc i} $\lambda 10830$ feature \citep{2023MNRAS.524.4447C}. 

In Figure~\ref{fig:spectra} (top panel), we compare the $+3$d spectrum of \obj\ with synthetic spectra from the double detonation models of \citet{2019ApJ...873...84P} in order to place rough limits on the amount of He that could be present in the SN ejecta. We chose models with He shell masses of 0.01, 0.03 and 0.05 \msun. The \citet{2019ApJ...873...84P} models are computed at 0.25 day intervals. For comparison with \obj, we computed the time of $r$-band maximum from the synthetic model light curves, and selected the epoch that is closest to the phase of $+3$d for the observed spectrum. Models with a WD mass of $1.0$ \msol were chosen for the comparison, since their predicted peak luminosity is closest to that observed for \obj. The strength of the Ti~{\sc ii} trough observed in \obj\ is roughly consistent with the 0.03 \msol He shell model. Models with He shell masses of 0.05 \msol and higher are much redder than \obj\ due to significant UV line blanketing at bluer wavelengths. On the other hand, models with higher WD masses of 1.1 and 1.2 \msol are significantly brighter at bluer wavelengths, and show a poor match to the continuum shape of the observed $+3$d spectrum of \obj. 

We note that for surface radioactivity scenarios such as $^{56}$Ni mixing and He shell detonation, the specific composition of radioactive isotopes can significantly affect the observational signatures. \citet{2021MNRAS.502.3533M} explored the effects of different He shell compositions on the photometric and spectroscopic properties and found it can have a big impact. For example, models with a post-explosion He shell composition that is dominated by IMEs such as $^{32}$S and $^{36}$Ar are considerably bluer around peak compared to models with shell composition dominated by iron group elements (IGEs) such as $^{44}$Ti, $^{48}$Cr, $^{52}$Fe and $^{56}$Ni \citep[eg.][]{2021MNRAS.502.3533M}.

\section{Discussion}\label{sec:discussion}

\subsection{Physical scenarios for the early excess}

An early flux excess in the light curve has been discovered in all 02es-like SNe Ia in recent times that have early time photometric coverage -- iPTF14atg \citep{2015Natur.521..328C}, iPTF14dpk \citep{2016ApJ...832...86C,2018ApJ...865..149J,2021ApJ...919..142B}, \jhr \citep{2017Natur.550...80J}, \yvq \citep{2020ApJ...898...56M,2021ApJ...919..142B} and \obj\ (this work). This is likely a generic feature of this subclass, thus serving as an effective tool to investigate the progenitor scenario and explosion mechanism. In Section~\ref{subsec:mosfit}, we fit the early excess in the light curve of \obj\ with a CSM interaction model using \mosfit. The model yields a CSM mass of $M_{\rm CSM} \approx 0.05$ \msol with an inner radius $R_0 \approx 3 \times 10^{14}\,$cm (Table~\ref{tab:params}). We now consider alternative physical mechanisms that could plausibly explain the early excess instead. 

\begin{figure*}
    %\hspace*{-0.8cm}
    \centering
    \includegraphics[width=0.8\linewidth]{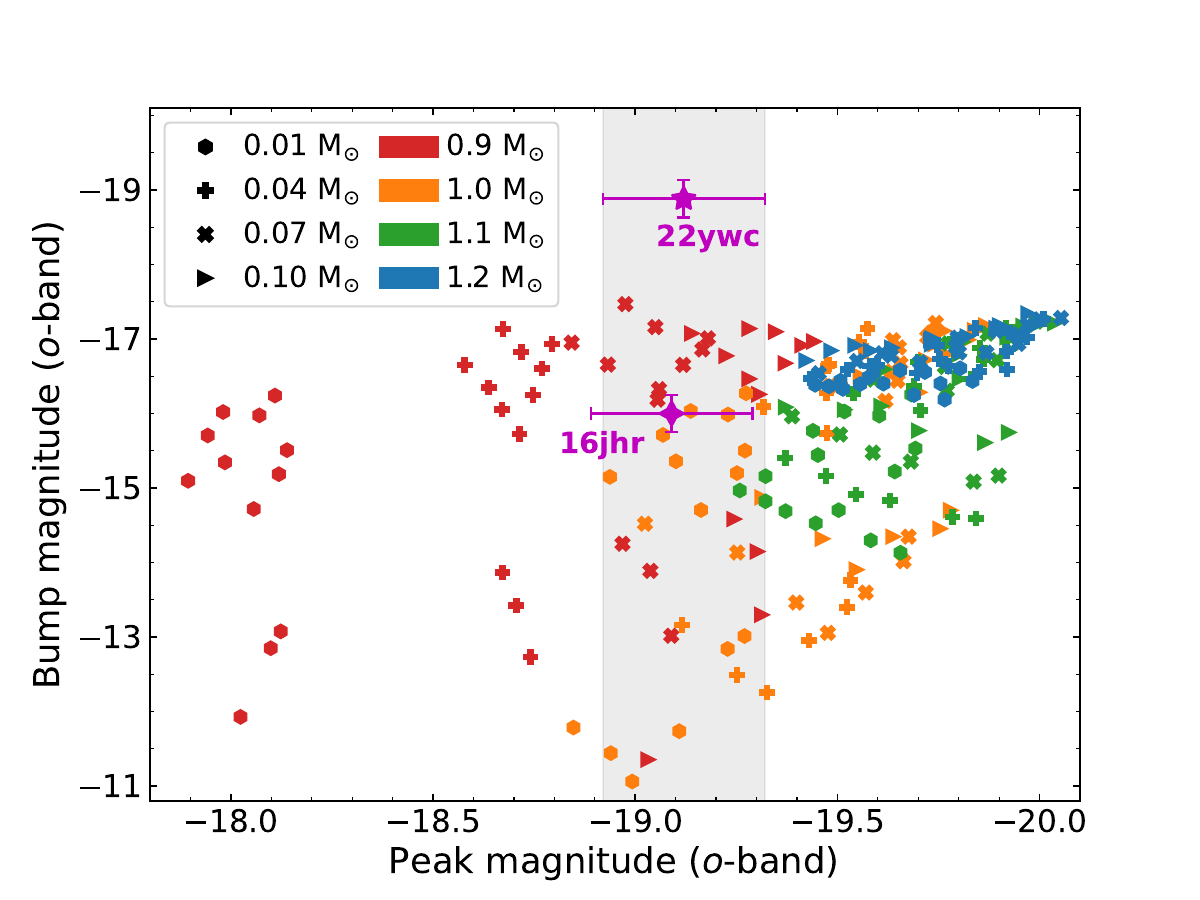}
    \caption{Predicted peak luminosity versus luminosity for the bump or early excess in the double detonation models of \citet{2021MNRAS.502.3533M} computed for $o$-band. The colors represent the WD masses and symbols represent the He shell mass. The shaded region highlights the models with a peak luminosity that is consistent with that of \obj. The peak magnitude of \obj\ and the associated uncertainty were estimated using the method described in Section~\ref{subsec:lc}.}
    \label{fig:bump_peak_mag}
\end{figure*}

At an absolute magnitude of $\sim -19$, the early excess for \obj\ is comparable in luminosity to the main peak. Given we require a $^{56}$Ni mass of $M_{\rm Ni} \approx 0.7$ \msol for the main peak, a comparable and therefore very large amount of surface $^{56}$Ni due to mixing or an extremely thick He shell would have to be invoked to explain the early excess in terms of surface radioactivity \citep{2019ApJ...873...84P,2020A&A...634A..37M,2021MNRAS.502.3533M}. The ATLAS $(c-o)$ color (Figure~\ref{fig:csmni}) and spectroscopic observations (Figure~\ref{fig:spectra}) do not show evidence of heavy line-blanketing effects at bluer wavelengths that would be expected from a large amount of radioactive material in the surface layers, thereby disfavoring these two scenarios. To illustrate this further, Figure~\ref{fig:bump_peak_mag} shows the predicted bump or excess luminosity versus the main peak luminosity for the suite of double detonation models in \citet{2021MNRAS.502.3533M}. We use the \citet{2021MNRAS.502.3533M} models to compute luminosities in the observer-frame $o$-band for a direct comparison with \obj. As expected, \obj\ stands out as an obvious outlier.

Next, we consider the ejecta-companion interaction scenario wherein SN ejecta colliding with a non-degenerate companion in the SD regime is shock-heated and drives a blue, UV-bright early excess \citep{2010ApJ...708.1025K}. The detection of an early excess in this scenario is sensitive to line-of-sight effects and is predicted to be observable only in $\sim 10\%$ of the events \citep{2010ApJ...708.1025K}. Apart from the viewing angle, detection of the early excess also depends on companion mass, binary separation and survey cadence and sensitivity \citep{2021ApJ...923..167W}. Although the fraction of events with a detectable early excess in this scenario may be higher than originally suggested, it nonetheless has a strong viewing angle dependence. The propensity for an early excess in Ia-02es suggests this feature may well be ubiquitous, and thus makes ejecta-companion interaction a less compelling scenario for this subclass in general. For the specific case of \obj, we examine the scaling relations in \citet{2010ApJ...708.1025K} for the luminosity from ejecta-companion interaction, $L \propto aE_{\rm k}^{7/8}M_{\rm ej}^{-7/8}$, where $a$ is the binary separation. Using a similar line of reasoning as \citet{2016ApJ...832...86C}, we estimate that the extraordinarily luminous early excess of \obj\ in the $c, o$ bands would require a kinetic energy $E_{\rm k} \sim 2 \times 10^{52}$ erg. This is an order of magnitude higher than the explosion energy expected from the disruption of a white dwarf \citep[eg.][]{2019NatAs...3..706J}, and also incompatible with the estimated kinetic energy from our \mosfit\ model of $E_{\rm k} \sim 1.3 \times 10^{51}$ erg. Given the constraints on ejecta velocity from spectra, and considering a reasonable range for WD progenitor masses, we conclude that the inferred kinetic energy for ejecta-companion interaction is too high and thus also rule out this scenario for the early excess observed in \obj.

\subsection{Implications for the progenitor scenario}

The ubiquity of early excess features in 02es-like SNe Ia would suggest a physical mechanism that is more isotropic in nature \citep{2021ApJ...919..142B}, as opposed to ejecta-companion interaction. It can be challenging to make a conclusive assessment for the scenario producing the early excess, even when high-cadence multi-band observations are available \citep{2017MNRAS.472.2787N}. The remarkably luminous early excess of \obj\ allows us to rule out surface radioactivity scenarios such as $^{56}$Ni clumps in the outer ejecta \citep{2020A&A...634A..37M} and thick He shell detonations \citep{2019ApJ...873...84P}, and also ejecta-companion interaction \citep{2010ApJ...708.1025K}. 

\citet{2015Natur.521..328C} invoked ejecta-companion interaction for the UV-bright early excess of iPTF14atg. However, only a specific optically thin CSM configuration was ruled out in that study \citep{2016MNRAS.459.4428K,2017hsn..book..317T}. Additionally, \citet{2016MNRAS.459.4428K} had difficulty reconciling the overall properties of iPTF14atg in terms of the SD scenario required for companion interaction, and instead favored a DD scenario involving a violent merger of sub-\mch WDs \citep{2013ApJ...770L...8P}. A violent merger scenario was also favored for 02es-like events \es \citep{2012ApJ...751..142G}, SN 2010lp \citep{2013ApJ...778L..18K} and PTF10ops \citep{2011MNRAS.418..747M}. The DD scenario for 02es-like events is strengthened by the detection of [O~{\sc i}] emission in nebular spectra of SN 2010lp \citep{2013ApJ...775L..43T} and iPTF14atg \citep{2016MNRAS.459.4428K}. The strong [Ca~{\sc ii}] emission in the nebular spectra of \yvq led \citet{2020ApJ...900L..27S} to favor a double detonation scenario, wherein an ignition in the He shell can cause a subsequent detonation in the carbon-oxygen core of a sub-\mch\ WD \citep{2010A&A...514A..53F,2010ApJ...714L..52S}. However, \citet{2021ApJ...914...50T} note that the combination of peculiar properties of \yvq including low luminosity, high velocity and lack of [O~{\sc i}] emission in nebular spectra are not easily reconciled with a single explosion model. A double detonation scenario was also invoked for \jhr by \citet{2017Natur.550...80J}, where the products of He shell burning account for the early excess in the light curve and the prominent Ti~{\sc ii} trough in the spectra. \citet{2017Natur.550...80J} suggest a model with a He shell mass of $\sim 0.05$ \msol and a primary WD mass in the range $1.28-1.38$ \msun, consistent with our $M_{\rm ej}$ estimate for \obj.

We thus favor CSM interaction as the likely scenario to explain this feature for \obj, and potentially for 02es-like events in general by extension. Our \mosfit\ model for \obj\ requires $\sim 0.05$ \msol of CSM, although its origin is unclear. A carbon-oxygen (CO) rich CSM can result in the DD scenario wherein the secondary WD is disrupted by the primary, forming a centrifugally supported disk \citep{2007MNRAS.380..933Y,2012ApJ...748...35S}. Simulations of WD mergers indicate that disruption of the secondary can eject $\sim 10^{-4}-10^{-2}$ \msol of material in the form of tidal tails \citep{2013ApJ...772....1R,2014MNRAS.438...14D}. The CSM distance in this case would depend on the time lag between ejection of the tidal tails and eventual explosion of the merged system. The simulations by \citet{2013ApJ...772....1R} suggest CSM distances of $\sim 10^{13}-10^{14}$ cm for a time lag of $\sim 10^6$ seconds ($\sim 10$ days).
This scenario involving a white dwarf exploding within a $\sim 10^{-2}-10^{-1}$ \msol CO-rich envelope has also been favored recently for 03fg-like SNe Ia \citep[eg.][]{2021ApJ...922..205A,2023MNRAS.521.1897M}. Another possibility is the CSM originating in the SD scenario involving a WD + He star binary \citep{2016A&A...589A..43N}.

Alternatively, the accretion disk formed following the disruption of the secondary WD could launch wind-driven material during the viscous phase of the merger along the polar directions, termed disk-originated matter or DOM \citep{2015MNRAS.447.2803L,2017MNRAS.470.2510L}. This model predicts $\sim 10^{-2}-10^{-1}$ \msol of DOM driven at $\sim 5000$ \kms\ \citep{2019ApJ...872L...7L}. If the explosion occurs within a few hours to days of the merger, the ejecta-DOM interaction would produce a detectable early excess. Assuming $v_{\rm DOM} \sim 5000$ \kms\ and $v_{\rm ejecta} \sim 13000$ \kms\ (from the \mosfit\ model), the estimated time lag between merger and explosion is $\sim 3$ days. The non-spherical configuration of the DOM could imply a favorable viewing angle along the polar axis for a small fraction of 02es-like events that would exhibit a pronounced early excess like \obj. Although this scenario presents a unified picture that is appealing, a larger sample and detailed modelling will be required to ascertain if it can reproduce the range of luminosity, colors and timescales for the early excess observed in 02es-like SNe Ia.

The overall photometric and spectroscopic properties of \obj\ (except for the luminous early excess) such as peak luminosity and decline rate (Figure~\ref{fig:lc_comp}), and the prominent Ti~{\sc ii} trough in the spectra (Figure~\ref{fig:spectra}) establish a connection with 02es-like SNe Ia in general, and in particular with hybrid 02es-like events such as \bt \citep{2010ApJ...708.1748F}, SN 2007cq \citep{2010ApJS..190..418G} and \jhr \citep{2017Natur.550...80J}. The CSM/DOM interaction scenario would require the secondary WD to be completely disrupted and accreted during the merger process \citep{2019ApJ...872L...7L}. This would disfavor a classic double detonation scenario for \obj\ since the secondary WD is expected to be intact when the primary explodes and is assumed to survive \citep{2013ApJ...770L...8P}, although recent simulations by \citet{2022MNRAS.517.5260P} investigate an interesting scenario where the secondary WD also explodes.

The evidence for He ash in the spectra of \obj\ could be plausibly explained by a merger involving a primary CO WD and a secondary He or HeCO WD \citep{2021ApJ...914...50T}, however detailed numerical simulations will be needed to confront this model with observations.

\subsection{Ia-02es and Ia-03fg: closely connected?}

Although disparate in terms of their peak luminosity, we note that 03fg-like and 02es-like SNe Ia share some traits that might suggest a common origin. Similar to 02es-like SNe Ia, recent 03fg-like events discovered at early times also display an early excess feature that has been attributed to ejecta-CSM interaction \citep{2021ApJ...923L...8J,2023MNRAS.521.1162D,2023ApJ...943L..20S}, suggesting this could be a unifying characteristic. Other similarities include weak or absent secondary $i$-band maxima \citep{2021ApJ...922..205A,2021ApJ...919..142B}, ejecta velocities that are typically lower than normal SNe Ia (although there are exceptions), and detection of [O~{\sc i}] emission in nebular spectra in some members of both sub-classes \citep{2013ApJ...775L..43T,2016MNRAS.459.4428K,2019MNRAS.488.5473T,2023MNRAS.521.1162D}. Early and late-time observations of a statistically large sample of 02es- and 03fg-like SNe Ia in the local volume will enable an in-depth investigation into this plausible connection.

\section{Conclusions}\label{sec:conclusion}

We have presented observations and analysis of the hybrid 02es-like Ia \obj, that shows a striking early excess in its light curve with no precedent in the literature. The remarkable luminosity of the early excess enables us to rule out surface radioactivity and ejecta-companion interaction scenarios. CSM interaction is favored as the likely scenario to power the early excess for \obj, and potentially for the subclass of Ia-02es in general. Our \mosfit\ model indicates a CSM mass of $\sim 0.05$ \msol at a distance of $\sim 3 \times 10^{14}$ cm from the WD. The origin of the CSM may be explained by ejection of tidal tails during a WD merger \citep{2013ApJ...772....1R}, or disk-originated matter (DOM) launched in bipolar directions from the accretion disk following the disruption of the secondary WD \citep{2019ApJ...872L...7L}. An asymmetric CSM configuration may be realised in either scenario.

The lack of multi-wavelength photometric observations (eg. UV and radio) for \obj\, and concurrent spectroscopic observations during the early excess limit us from placing stringent constraints on the CSM properties. Synchrotron radiation from relativistic electrons produced from ejecta-CSM interaction is expected to give rise to a luminous excess in the high frequency radio regime ($\sim 250$ GHz), that could last for a few days past explosion \citep{2023MNRAS.tmp.2257H}.
A multi-wavelength observational campaign will thus be imperative for the next \obj-like event to better understand the nature of these explosions.
Nebular phase observations for \obj\ (and in general for Ia-02es and Ia-03fg) would also be important for further constraining the nature of the progenitor \citep[eg.][]{2023arXiv230611788S}.

\section*{Data Availability}

Photometric data of \obj\ is publicly available through the ATLAS forced photometry server \citep{2021TNSAN...7....1S}. The stacked ATLAS photometry presented in Figure~\ref{fig:lc_comp} is available in the online version in machine-readable form. Spectroscopic data of \obj\ is available on the Weizmann Interactive Supernova Data Repository \citep[WISeREP;][]{2012PASP..124..668Y}.

%\begin{acknowledgments}
\vspace{2em}
ATLAS is primarily funded through NASA grants NN12AR55G, 80NSSC18K0284, and 80NSSC18K1575. The ATLAS science products are provided by the University of Hawaii, QUB, STScI, SAAO and Millennium Institute of Astrophysics in Chile. This work is based on observations collected at the European Organisation for Astronomical Research in the Southern Hemisphere, Chile, as part of ePESSTO+ (the advanced Public ESO Spectroscopic Survey for Transient Objects Survey). ePESSTO+ observations were obtained under ESO program ID 108.220C (PI: Inserra). SS thanks G. Dimitriadis and C. Frohmaier for useful discussions. MN, SS and XS are supported by the European Research Council (ERC) under the European Union's Horizon 2020 research and innovation programme (grant agreement No.~948381) and by UK Space Agency Grant No.~ST/Y000692/1. SS, SAS and SJS acknowledge funding from STFC Grants ST/X006506/1 and ST/T000198/1. LG and TEMB acknowledge support from Unidad de Excelencia Mar\'ia de Maeztu CEX2020-001058-M, from Centro Superior de Investigaciones Cient\'ificas (CSIC) under the PIE project 20215AT016, and from the Spanish Ministerio de Ciencia e Innovaci\'on (MCIN) and the Agencia Estatal de Investigaci\'on (AEI) 10.13039/501100011033 under the PID2020-115253GA-I00 HOSTFLOWS project. LG also acknowledges support from the European Social Fund (ESF) ``Investing in your future'' under the 2019 Ram\'on y Cajal program RYC2019-027683-I. TEMB also acknowledges financial support from the 2021 Juan de la Cierva program FJC2021-047124-I. This work was funded by ANID, Millennium Science Initiative, ICN12\_009. GP is supported by the Millennium Science Initiative through grant IC120009, awarded to The Millennium Institute of Astrophysics (MAS). MRM acknowledges a Warwick Astrophysics prize post-doctoral fellowship made possible thanks to a generous philanthropic donation. We thank the anonymous referee for constructive comments on the manuscript.

%\end{acknowledgments}


%\vspace{5mm}
\facilities{ATLAS, NTT}


\software{astropy \citep{2013A&A...558A..33A,2018AJ....156..123A}, MOSFiT \citep{2018ApJS..236....6G}}



%\appendix


\bibliography{ref}
\bibliographystyle{aasjournal}



\end{document}